\documentclass[onecolumn,10pt]{article}
\usepackage{newtxtext}
\usepackage{titling}
\usepackage{tabularx}
\usepackage{ragged2e} 

\usepackage{hhline}
\usepackage{graphicx,graphics,color}
\usepackage{amsmath,amssymb,cite,enumitem,mathtools}
\usepackage[]{hyperref}
\usepackage[scriptsize]{subfigure}
\usepackage[small]{caption}
\usepackage{bbding}
\usepackage{lineno}
\usepackage{bm}
\usepackage{calligra}
\usepackage{authblk}
\usepackage{etoc}
\usepackage{resizegather}
\usepackage{algorithm}
\usepackage{algorithmic}
\usepackage{lscape}
\usepackage{float}

\usepackage{vmargin}
\setmarginsrb  { 2.5cm}  
                        { 2.5cm}  
                        { 2.5cm}  
                        { 2.5cm}  
                        { 0cm }  
                        { 1.2cm }  
                        {  0cm }  
                        { 1.0cm }  
                        
\usepackage{setspace}
\usepackage{mathpazo}

\usepackage{titlesec}
\titleformat{\section}
  {\normalfont\large\bfseries}
  {\thesection}{1em}{}

\usepackage{multirow}
\usepackage{wasysym,hhline,booktabs}

\newcommand*\patchAmsMathEnvironmentForLineno[1]{%
  \expandafter\let\csname old#1\expandafter\endcsname\csname #1\endcsname
  \expandafter\let\csname oldend#1\expandafter\endcsname\csname end#1\endcsname
  \renewenvironment{#1}%
     {\linenomath\csname old#1\endcsname}%
     {\csname oldend#1\endcsname\endlinenomath}}%
\newcommand*\patchBothAmsMathEnvironmentsForLineno[1]{%
  \patchAmsMathEnvironmentForLineno{#1}%
  \patchAmsMathEnvironmentForLineno{#1*}}%
\AtBeginDocument{%
\patchBothAmsMathEnvironmentsForLineno{equation}%
\patchBothAmsMathEnvironmentsForLineno{align}%
\patchBothAmsMathEnvironmentsForLineno{flalign}%
\patchBothAmsMathEnvironmentsForLineno{alignat}%
\patchBothAmsMathEnvironmentsForLineno{gather}%
\patchBothAmsMathEnvironmentsForLineno{multline}%
}

\usepackage{array} 
\usepackage{ifthen} 
\usepackage{ulem}
\usepackage[overload]{empheq}

\newcommand{\de}[1][]{\text{d}}
\newcommand{\F}[1][]{{}_2F_1}

\newcommand{\nn}[1][]{\text{NN}_\theta}
\newcommand{\NN}[1][]{\text{NN}}
\newcommand{\son}[1][]{\sigma_\text{on}}
\newcommand{\soff}[1][]{\sigma_\text{off}}
\newcommand{\dm}[1][]{d_\text{m}}
\newcommand{\dpr}[1][]{d_\text{p}}
\newcommand{\nm}[1][]{n_\text{m}}
\newcommand{\np}[1][]{n_\text{p}}

\newcommand{\ra}[1][]{\epsilon_a}
\newcommand{\rb}[1][]{\epsilon_b}

\hypersetup{
    colorlinks=true,     
    linkcolor=blue,     
    citecolor=blue,     
    filecolor=black,     
    urlcolor=blue       
}

\graphicspath{{Fig/}}

\title{Universal deterministic patterns in stochastic count data}
\author[1,2$^*$]{Zhixing Cao}
\author[2]{Yiling Wang}
\author[3$^*$]{Ramon Grima}
\affil[1]{\textit{Department of Chemical Engineering, Queen's University, Canada}}
\affil[2]{\textit{State Key Laboratory of Bioreactor Engineering} \protect\\\textit{East China University of Science and Technology, China}}
\affil[3]{\textit{School of Biological Sciences, University of Edinburgh, United Kingdom}}

\affil[$^*$]{email: \texttt{z.cao@queensu.ca} \& \texttt{ramon.grima@ed.ac.uk}}

\begin{document}

\maketitle

\section*{Abstract}
We report the existence of deterministic patterns in plots showing the relationship between the mean and the Fano factor (ratio of variance and mean) of stochastic count data. These patterns are found in a wide variety of datasets, including those from genomics, paper citations, commerce, ecology, disease outbreaks, and employment statistics. We develop a theory showing that the patterns naturally emerge when data sampled from discrete probability distributions is organised in matrix form. The theory precisely predicts the patterns and shows that they are a function of only one variable --- the sample size.

\newpage

\section*{Introduction} 

Measurements of discrete, stochastic count data are often presented in matrix form. A prominent example of this type of data comes from single-cell transcriptomic techniques, such as single-cell RNA sequencing (scRNA-seq \cite{tang2009mrna,zheng2017massively,hagemann2020single}) and multiplexed error-robust fluorescence in situ hybridization (MERFISH \cite{chen2015spatially,xia2019spatial,zhang2021spatially}), which enable the study of gene expression at the individual cell level. The data is organised as a matrix where the integer at the $i$-th row and $j$-th column is the transcript count for gene $i$ in cell $j$. The discreteness of the data naturally stems from the fact that we are counting molecule numbers and its stochastic nature is due to intrinsic and extrinsic sources of noise \cite{elowitz2002stochastic}. 

Other types of discrete and stochastic data that are presentable in an intuitive matrix form include those from paper citations, commerce, disease outbreaks and employment statistics which are available from various publicly available repositories. For example, for citation data, each row of the count matrix could correspond to a scientific paper published in say the year 2000 and each column to a time window such as a month beginning from January next year. The matrix entry at the $i$-th row and $j$-th column in the number of citations of the $i$-th paper from 2000 due to the new papers submitted in the $j$-th month starting from January 2001. Another common example of discrete stochastic count data is epidemiological. Here, the count matrix has rows corresponding to an infectious disease or condition and columns to a location; the matrix entry at the $i$-th row and $j$-th column is the number of people with disease $i$ in location $j$, counted over a certain fixed period of time. 

Our aim is to show that independent of the degree of stochasticity and the interpretation of the discrete stochastic data, non-trivial deterministic patterns emerge when the low-order statistics of the data are plotted in a natural way, and that these universal patterns can be fully explained by a simple theory. 

\section*{Deterministic patterns in single-cell transcriptomic data}

We start by focusing on single-cell transcriptomic data. Due to the considerable stochasticity in this data, it is common to report measures of the size of molecular noise \cite{kaern2005stochasticity,sanchez2013genetic}. A simple variability measure is the Fano factor --- the ratio of the variance and the mean of transcript counts, separately calculated for each gene using the data in each row of the matrix. The Fano factor ($\text{FF}$) equals one if the distribution of counts is Poissonian; it is greater (smaller) than 1 if the distribution of counts is wider (narrower) than a Poisson with the same mean. Cases of sub-Poisson expression ($\text{FF} < 1$,\cite{weidemann2023minimal}), Poisson expression ($\text{FF} = 1$,\cite{zenklusen2008single}) and super-Poisson expression ($\text{FF} > 1$,\cite{dar2012transcriptional}) have been reported in the literature.  

\begin{figure}[!t]
\includegraphics[width=\textwidth]{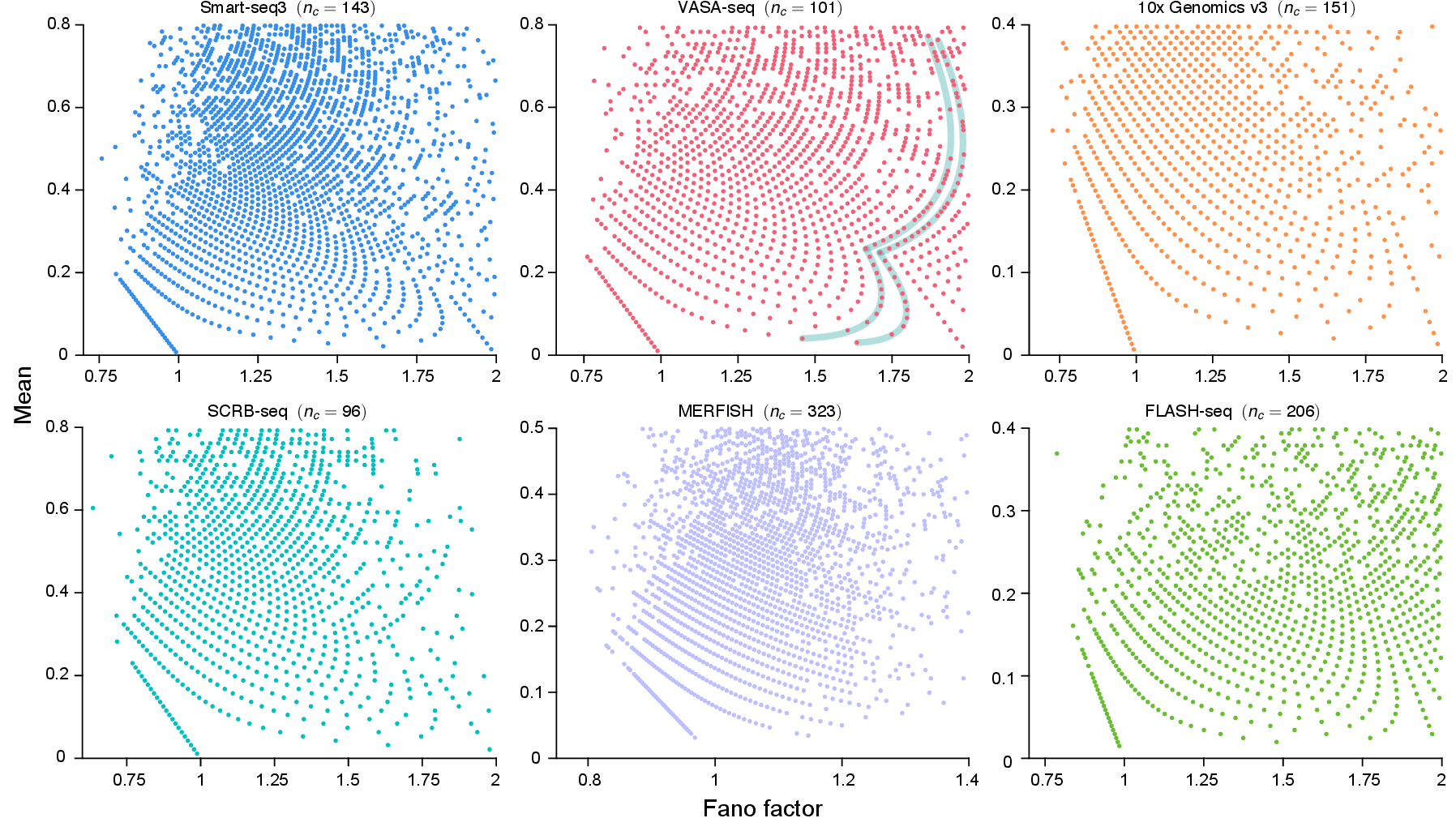}
\caption{Mean-Fano factor plots of six genomic datasets using various types of scRNA-seq protocols and MERFISH. Each plot shows a deterministic pattern; one pattern is highlighted for VASA-seq data. Each point represents the mean and Fano factor computed from the transcript counts of a gene in a finite number $n_c$ of cells (some genes may have the same coordinates). Information on each dataset can be found in \ref{datainfo}.}\label{fig1}
\end{figure}

To understand how the Fano factor might depend on the mean expression, the type of measurement technique and the cell types, we computed plots of the mean versus the Fano factor using yeast, mouse and human data collected using MERFISH and 5 different types of scRNA-seq protocols. The plots are shown in Fig. \ref{fig1}. Note that each point in this plot represents a gene and we have specifically focused on the region where the mean transcript count is less than 1. 

Contrary to our expectations, the plots are not particularly different from each other and in all cases they show the existence of deterministic patterns. This is particularly surprising given the stochastic nature of gene expression \cite{elowitz2002stochastic} and the large differences between sequencing protocols \cite{ziegenhain2017comparative}. All together, the evidence suggests that the patterns are not an artefact of the experimental technique and that they may have an underlying common origin due to fundamental mathematical, physical, chemical or biological principles. 

\begin{figure}
\includegraphics[width=0.95\textwidth]{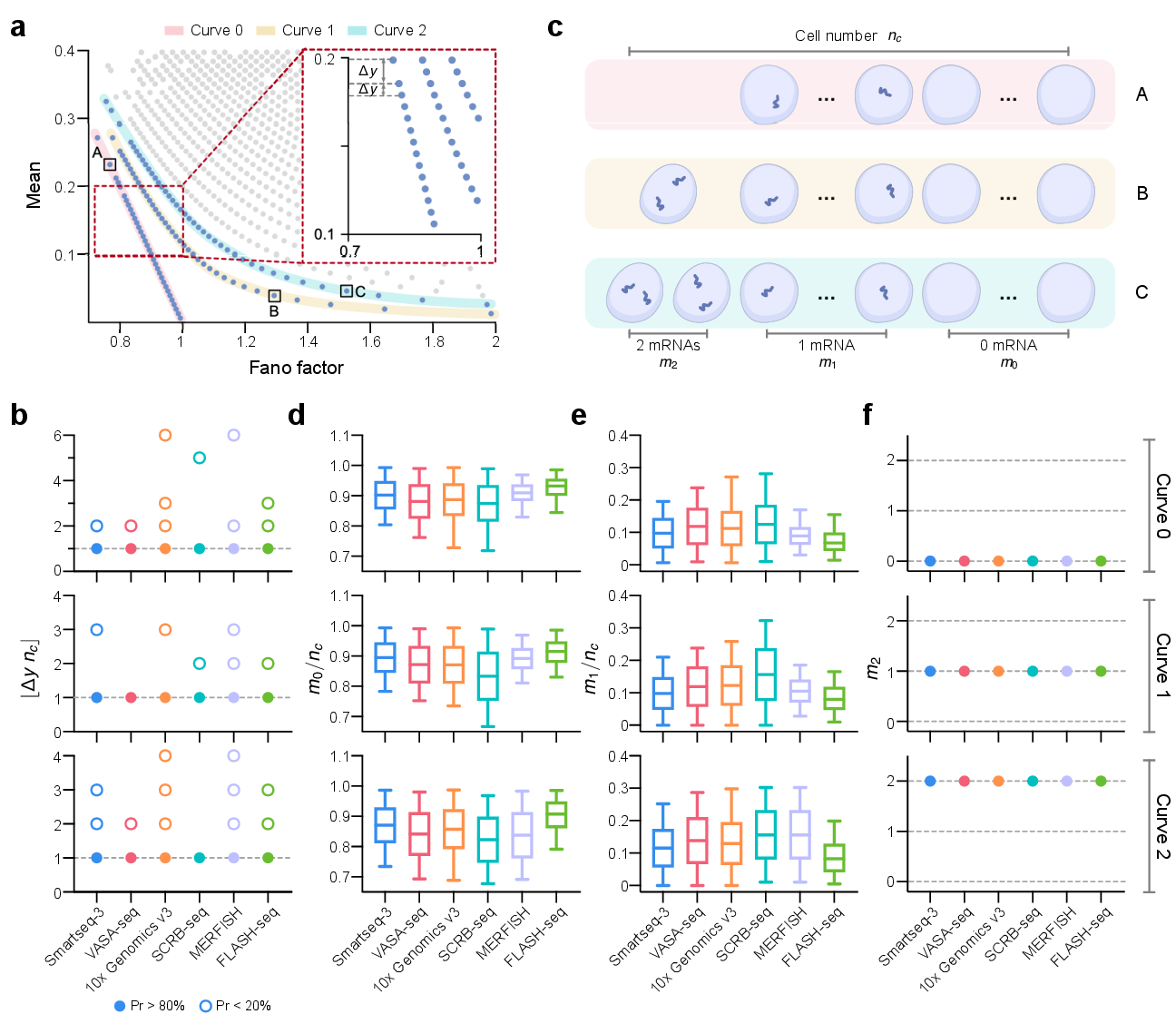}
\caption{Analysis of the patterns in the mean-Fano factor plot of single-cell transcriptomic data. (a) Points in the mean-Fano factor plot generated using the 10x Genomics v3 technology are arranged on distinct curves. We highlight three of them and call them Curves 0, 1 and 2. The inset zooms in on the curves and shows the existence of a periodic distance between the points on the same curve. The vertical distance between two successive points on the same curve is denoted by $\Delta y$. (b) For all six datasets in Fig. \ref{fig1}, $\Delta y$ (computed from Curves 0-2) is a multiple of $1/n_c$ where $n_c$ is the number of cells used to compute the mean and the Fano factor of transcript number fluctuations. Note that in $80\%$ of cases, $\Delta y = 1/n_c$ (solid circles). (c) Variation in the number of transcripts per cell for three genes A, B and C on Curves 0, 1 and 2 respectively in (a). For gene A, cells only have 0 or 1 transcript. For gene B, cells have 0-2 transcripts but only one cell has 2 transcripts. For gene C, cells have 0-2 transcripts but only two cells have 2 transcripts. Note that $m_i$ is the number of cells with exactly $i$ transcripts. (d) Fraction of cells with exactly zero transcripts for genes on Curves 0 (top), 1 (middle) and 2 (bottom) for all six datasets. (e) Same as (d) but showing fraction of cells with exactly one transcript. Table \ref{tabS1} summarises the median values in (d) and (e). (f) Number of cells with exactly two transcripts for genes on Curves 0 (top), 1 (middle) and 2 (bottom) for all six datasets.}\label{fig2}
\end{figure}

An important observation is that the points in the mean-Fano factor plot appear to fall on a set of monotonically decreasing curves from left to right. In Fig. \ref{fig2}a we highlight three of these curves in the plot generated using the 10x Genomics v3 platform and refer to them as Curves 0, 1 and 2. At first glance, it appears that the vertical distance between two successive points on the same curve, $\Delta y$, almost always takes the same value (inset of Fig. \ref{fig2}a). This quantization is verified in Fig. \ref{fig2}b, where we show that for all points on Curves 0, 1 and 2 in the six datasets in Fig. \ref{fig1}, $\Delta y$ is a multiple of $1/n_c$ where $n_c$ is the number of cells in each dataset (number of columns in the count matrix). Furthermore, in $80\%$ of the cases, $\Delta y = 1/n_c$. This suggests that $\Delta y$ is a multiple of $1/n_c$ because of missing data points, i.e. either the gene corresponding to this point does not exist or else it was not detected due to imperfect capture of a proportion of transcripts.

Next we investigated whether there are any obvious statistical differences between the expression of genes on Curves 0, 1 and 2 in the 10x Genomics v3 dataset. In Fig.  \ref{fig2}c we show that for a gene on Curve 0 (denoted as gene A in Fig. \ref{fig2}a) each cell in the sample has only 0 or 1 transcripts; for a gene on Curve 1 (denoted as gene B in Fig. \ref{fig2}a) one cell has two transcripts and the rest of the cells in the sample have only 0 or 1 transcripts; for a gene on Curve 2 (denoted as gene C in Fig. \ref{fig2}a) two cells have two transcripts and the rest of the cells in the sample have only 0 or 1 transcripts. In Fig. \ref{fig2}d-f we show that these observations extend to all genes in Curves 0-2 in all six datasets. Namely, genes on different curves cannot be distinguished by the fraction of cells with 0 or 1 transcripts but only by the number of cells which have 2 transcripts. Armed with these observations, we now devise a simple theory that can predict the patterns in the mean-Fano factor plots. 

\section*{A simple theory can explain the pattern}

\begin{figure}[!hb]
\includegraphics[width=\textwidth]{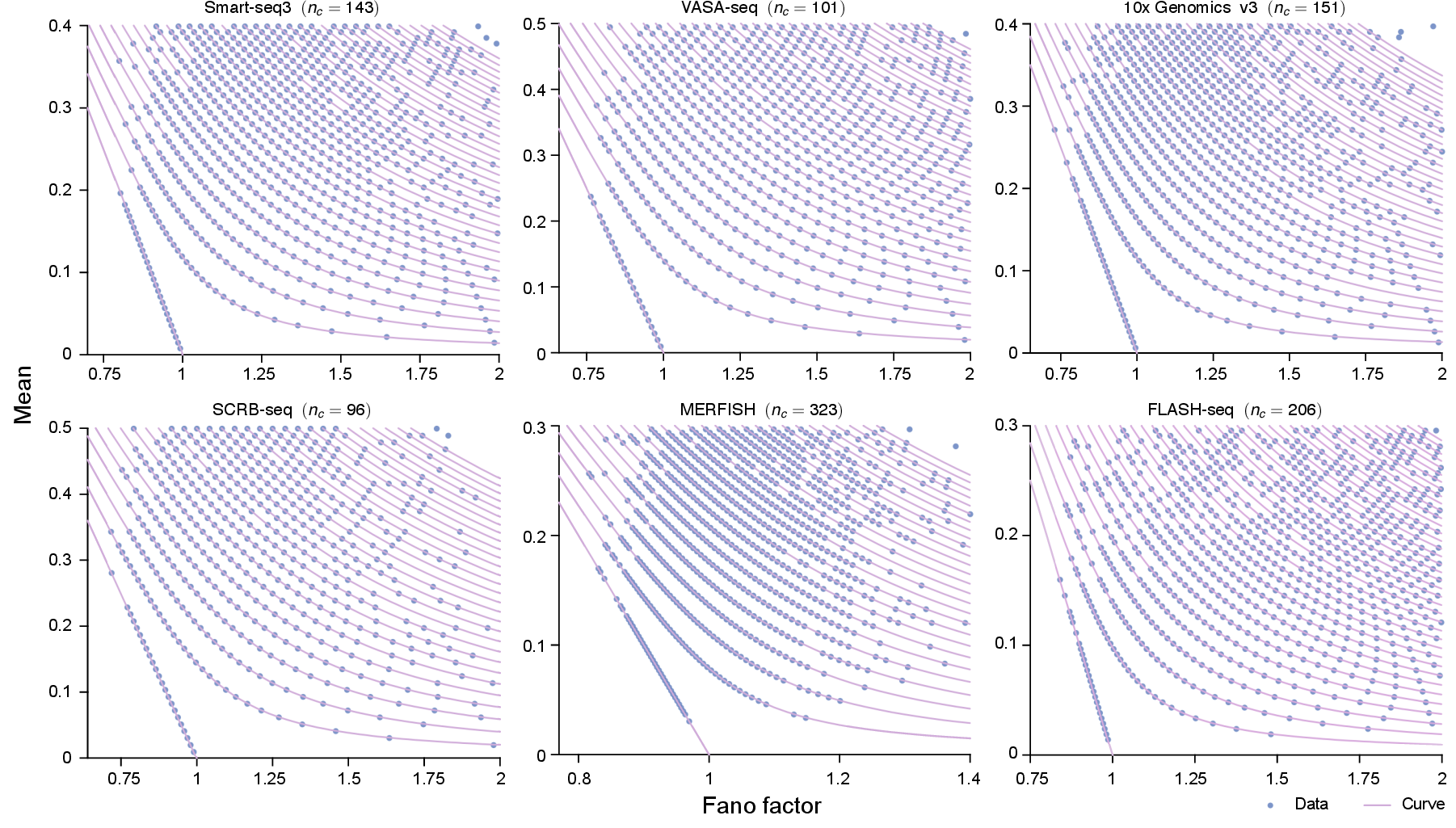}
\caption{Mean-Fano factor plots of the six sequencing datasets in Fig. \ref{fig1} and the theoretical predictions given by Eqs \eqref{curve_final}-\eqref{curve_final1}. The curves pass through all points in the dataset thus verifying the accuracy of the theory.}\label{sfig1}
\end{figure}

Consider genes sitting on Curve $k$ which can have only 0, 1 or 2 transcripts with the proviso that there are only $k$ cells in the sample which have exactly two transcripts. Let $n_c$ be the total number of cells. Then the expression data for a gene is given by the vector $\{ n_1, n_2, n_3, \cdots, n_{n_c-k}, 2, \cdots, 2 \}$ where $n_i$ is the number of transcripts in cell $i$. It then follows that the mean and the mean squared expression are given by:
\begin{equation}
    \label{meanexp}
    \langle n \rangle = \frac{\sum_{i=1}^{n_c-k} n_i + 2k}{n_c} = a + \frac{2k}{n_c}, 
    \quad \langle n^2 \rangle = \frac{\sum_{i=1}^{n_c-k} n_i^2 + 2^2 k}{n_c} = a + \frac{4k}{n_c},
\end{equation}
where $\sum_{i=1}^{n_c-k}n_i /n_c = \sum_{i=1}^{n_c-k}n_i^2 /n_c$ equals some fraction $a$; this is since $n_i = 0$ or $1$ for $1 \le i \le n_c - k$. Hence the Fano factor is given by
\begin{equation}
    \label{FFeq}
    {\rm{FF}}= \frac{ \langle n^2 \rangle -  \langle n \rangle^2}{\langle n \rangle} = 1 - \langle n \rangle + \frac{2k}{n_c \langle n \rangle}, \quad 1 \le k \le n_c,
\end{equation}
where we eliminated the parameter $a$ using Eq. \eqref{meanexp}. Solving for the mean, we finally obtain the equation of Curve $k$ in the mean-Fano factor plot:
\begin{align}\label{curve_final}
\langle n \rangle&=1 - \text{FF}, \quad k = 0  \\ \label{curve_final1}
\langle n \rangle&=\frac{1}{2}\left(1-\text{FF}+\sqrt{\frac{8k+n_c(\text{FF}-1)^2}{n_c}}\right), \quad 1 \le k \le n_c.
\end{align}
Note that genes on Curve 0 have a Bernoulli distribution of transcript counts because in each cell the count is either 0 or 1. Eqs. \eqref{curve_final}-\eqref{curve_final1} explain why the plotting of 10x Genomics data in Fig. \ref{fig2}a shows Curve 0 is a straight line while Curves 1 and 2 have non-zero curvature. It is straightforward to show that for any discrete distribution of transcript numbers, $\langle n \rangle \ge 1 - \text{FF}$ (\ref{nopoints}). This explains the existence of the empty bottom left triangle bounded from above by Curve 0 in Fig. \ref{fig2}a. In Fig. \ref{sfig1} we verify that the theory is successful in predicting the curves in the mean-Fano factor plots of all six types of single-cell transcriptomic data shown in Fig. \ref{fig1}.

\begin{figure}[!b]
\centering
\includegraphics[width=\textwidth]{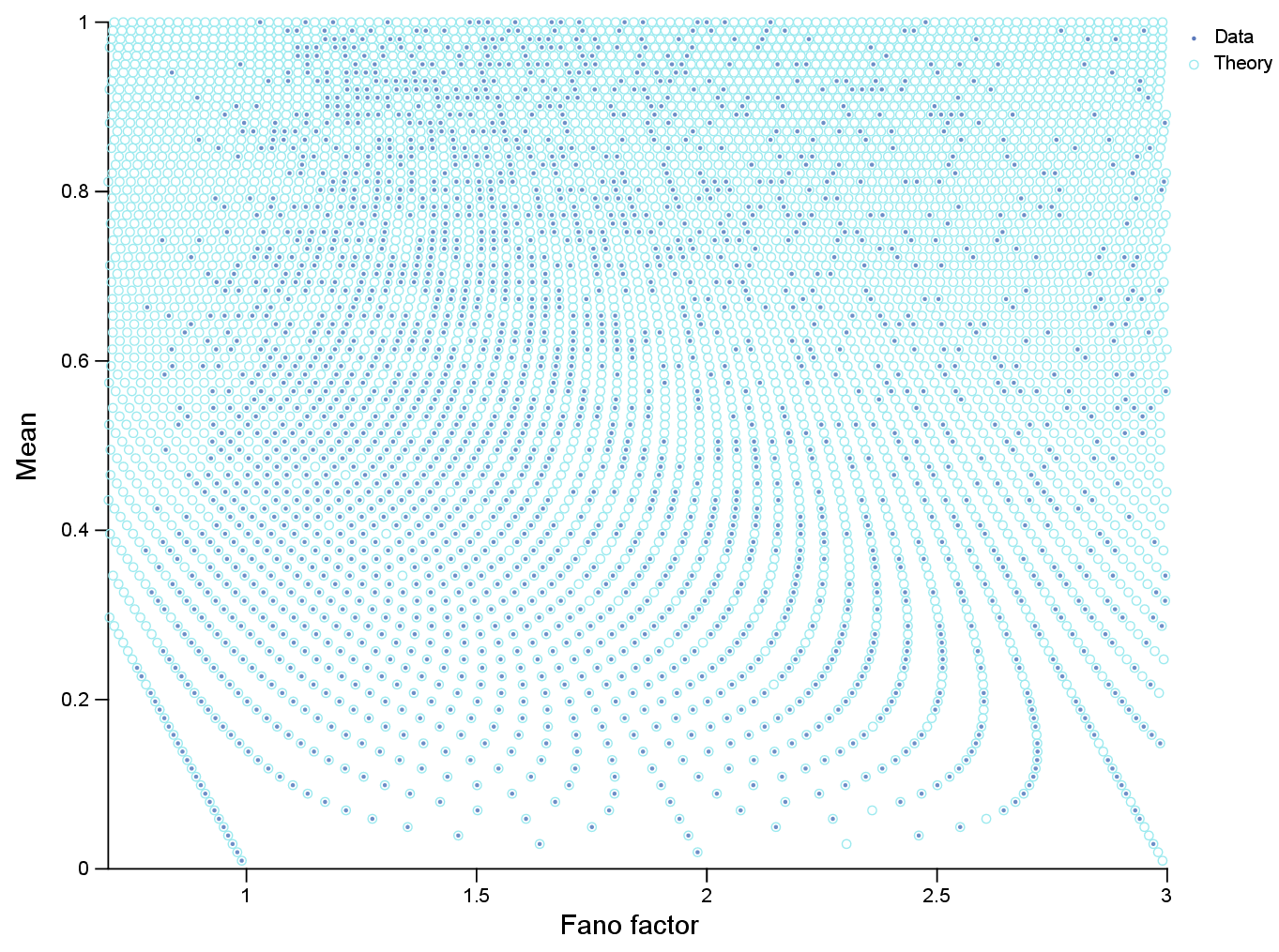}
\caption{Theory predicts the patterns in the mean-Fano factor plot of VASA-seq data. The dots are calculated from the data (each represents a gene) and the open circles have x-coordinate given by Eq. \eqref{FFeq} and y-coordinate given by $\langle n \rangle = i/n_c$ where $n_c$ is the sample size of 101 cells and $i$ is a positive integer. Note that some open circles lack corresponding dots due to missing data points. This absence occurs either because the gene corresponding to the point does not exist or it was not detected because typically only a small fraction of the transcriptome of each cell is captured by sequencing methods \cite{ziegenhain2017comparative}.}\label{fig3}
\end{figure}

The theory also provides a simple explanation of the quantization along the vertical axis shown in Fig. \ref{fig2}b. Recall that the mean expression $\langle n \rangle$ of a gene on Curve $k$ is given by Eq. \eqref{meanexp}. Since the numerator of the equation is an integer it follows that the mean transcript count of genes on Curve $k$ can only differ by an integer multiple of $1/n_c$. Hence our hypothesis is that by choosing the y-coordinate position of a point to be a multiple of $1/n_c$ and its x-coordinate to be given by Eq. \eqref{FFeq}, we can reproduce the full point pattern. In Fig. \ref{fig3} we verify that this recipe precisely predicts the intriguing pattern in the mean-Fano factor plot of the VASA-seq dataset in Fig. \ref{fig1}. In particular, the position of the data points (shown by the dark blue dots) coincides with the blue open circles predicted by the theory. The success of the simple theory is in large part because its main results agree with those from a more general theory where we consider genes sitting on curves which can have 0, 1, $\cdots$, $N$ transcripts with the proviso that there are exactly $m_i$ cells in the sample with $i \ge 2$ transcripts (\ref{gencurve}).

\begin{figure}[!t]
\includegraphics[width=\textwidth]{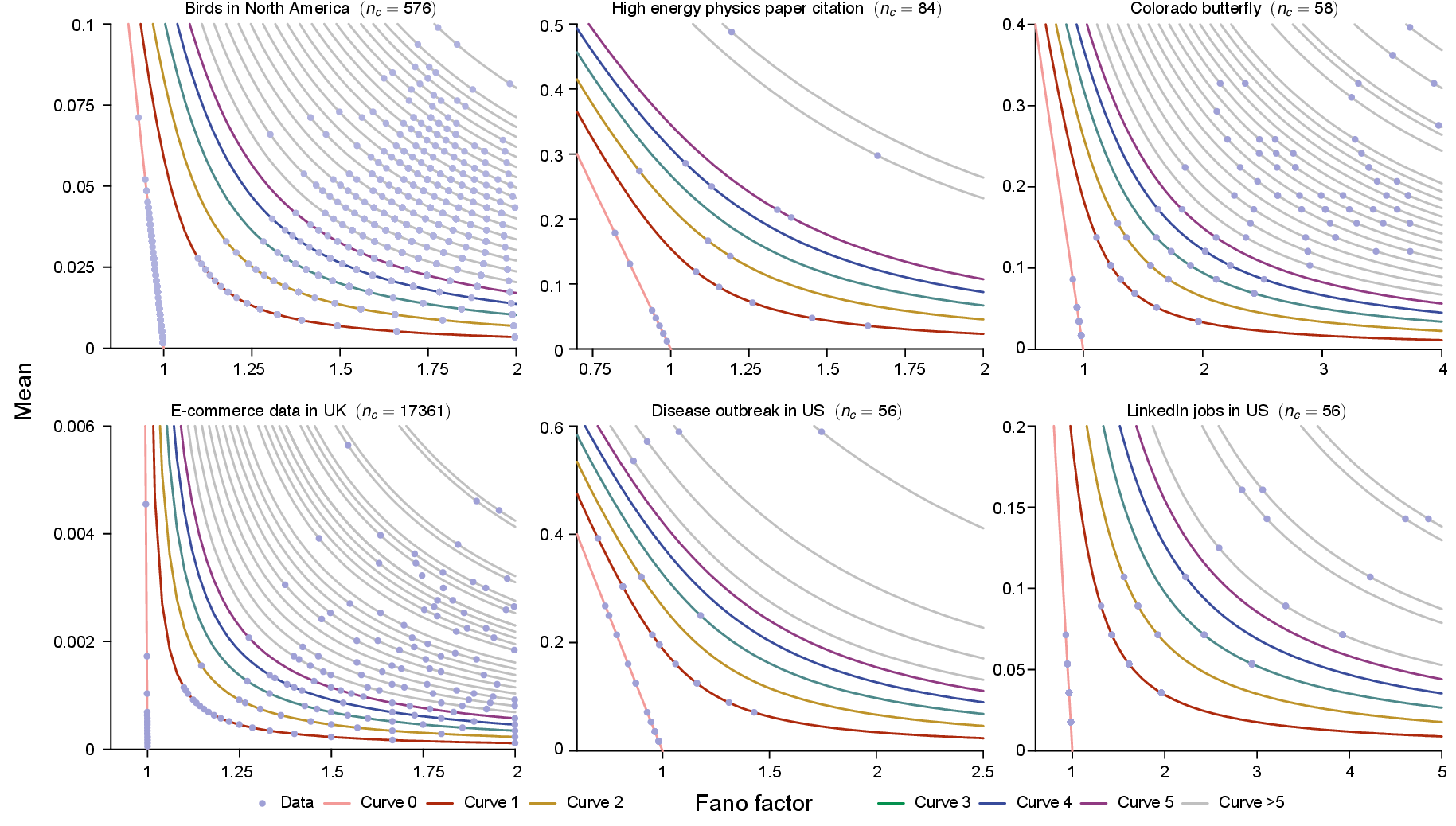}
\caption{mean-Fano factor plots of six different types of discrete and stochastic datasets which are not related to single-cell transcriptomics. Further information on each dataset can be found in \ref{datainfo}. The theoretical equation of Curve $k$ is given by Eqs. \eqref{curve_final}-\eqref{curve_final1}. The variable $n_c$ is the number of columns in the count matrix. The number of points in the plot is equal to the number of rows in the count matrix.
In all cases, the theoretical curves exactly pass through the points in the plot thus verifying the theory's accuracy for diverse types of datasets. Note that for visual clarity, we only plot those theoretical curves with $k>5$ which explain the measured point positions.}\label{fig4}
\end{figure}

Finally to test the theory's predictions more broadly we computed the mean-Fano factor plots of six datasets that were not generated by single-cell transcriptomic techniques. Two of these datasets are taken from the field of ecology (counting of the number of birds and butterflies of various species observed at several locations in North America). The other four datasets use citation information of high-energy physics papers submitted to a preprint server, the number of purchased items of various types by customers in the UK, the number of people with certain diseases or conditions organised by location in the USA and employment statistics for various jobs also organised by location in the USA. Details for each dataset can be found in \ref{datainfo}. All these datasets were arranged in the form of a count matrix, akin to the gene$\times$cell count matrix used to previously analyse single-cell data. The coordinates of the $i$-th point in the mean-Fano factor plot are computed using the data entries in the $i$-th row vector of the count matrix.  In Fig. \ref{fig4} we show plots of the mean-Fano factor for the six datasets. In all cases we find that the theoretical curves given by Eqs. \eqref{curve_final}-\eqref{curve_final1} precisely pass through all the points in the plots.

\section*{Conclusion}

In this paper we have reported the existence of deterministic patterns in the mean-Fano factor plots summarising the statistics of data from single-cell transcriptomics, paper citations, commerce, ecology, disease outbreaks and employment. The patterns are only obvious for datasets with small sample sizes and specifically when one zooms in on the region where the mean is small. In the analysis of single-cell transcriptomic datasets, particularly those obtained from single-cell sequencing, data with a small mean is often removed from subsequent data analysis \cite{luecken2019current} which might explain why the patterns have not been reported before. Emergence of deterministic patterns from highly stochastic data seems counterintuitive and hence our first hypothesis was that this pattern is due to artefacts introduced by the experimental measurement techniques. However, our theory precisely predicts the patterns and shows that they are simply a mathematical consequence of the discreteness and stochasticity of the data which hence explains the ubiquitousness of the patterns in data collected using six different types of single-cell transcriptomic techniques and also in completely unrelated datasets. All together, the study shows that while the existence of naturally occurring patterns often propels us to think of the underlying biological, chemical or physical mechanisms that may be causing them, sometimes the answer is much simpler!     

\section*{Code and data availability}
The Julia code is deposited at \href{https://github.com/edwardcao3026/pattern}{Github}, and the data is uploaded to \href{https://doi.org/10.5281/zenodo.11213863}{Zenodo}.

\section*{Competing interests}
There are no competing interests declared.

\section*{Acknowledgements}

Z.C. acknowledges support from NSFC Grant (62073137), Shanghai Action Plan for Technological Innovation Grants (22ZR1415300, 22511104000, 23S41900500), the Natural Science and Engineering Research Council of Canada's (NSERC's) Discovery Grant (RGPIN-2024-06015) and the Shanghai Center of Biomedicine Development. R.G. acknowledges support from a Leverhulme Trust grant (RPG-2020-327).  The authors would like to thank James Holehouse and Augustinas Sukys for useful feedback. 

\renewcommand{\emph}[1]{\textit{#1}}
\bibliographystyle{naturemag.bst}

\renewcommand{\emph}{\textit}


\clearpage
 \makeatletter \newcommand{\emptyauthor}{\renewcommand{\author}[1]{} \renewcommand{\@author}{}} \makeatother
\emptyauthor 

\title{Supplementary Information}

\date{Zhixing Cao, Yiling Wang, Ramon Grima\\[2em] \today}

\maketitle

\clearpage

\clearpage
\setcounter{section}{0}
\renewcommand{\thesection}{Supplementary Note \arabic{section}}
\setcounter{equation}{0}
\setcounter{figure}{0}
\setcounter{table}{0}
\renewcommand{\theequation}{S\arabic{equation}}
\renewcommand{\thetable}{S\arabic{table}}
\renewcommand{\thefigure}{S\arabic{figure}}

\section{}\label{datainfo}
In this section, we describe the collection and preprocessing of count data analyzed in the main text.
\begin{itemize}
\item 
\textbf{Smartseq-3:} Unique molecular identifier (UMI) count data (\texttt{Smartseq3.HEK.cleanup.UMIcounts.\\txt}) and the annotation file (\texttt{Smartseq3.HEK.cleanup.sample\_annotation.txt}) from \href{https://www.ebi.ac.uk/biostudies/arrayexpress/studies/E-MTAB-8735}{E-MTAB-8735} provided in Ref. \cite{hagemann2020single} were accessed. Based on the annotation file, 143 HEK293T cells that underwent template switch oligo (TSO) and forward primer addition were specifically chosen to compute the count data matrix.  
\item
\textbf{VASA-seq:} The count data for unique fragment identifiers (UFIs) (\texttt{GSM5369497\_E6.5-1\_i1\_total.\\UFICounts.tsv}) were obtained from \href{https://www.ncbi.nlm.nih.gov/geo/query/acc.cgi?acc=GSE176588}{GSE176588}, as described in Ref. \cite{salmen2022high}. This dataset comprises 101 E6.5 mouse embryo cells. 
\item
\textbf{10x Genomics:} The 10x Genomics Chemistry v3 sequencing data was obtained from the file \texttt{pbmc\_1k\\\_v3\_filtered\_feature\_bc\_matrix.h5} on the \href{https://cf.10xgenomics.com/samples/cell-exp/3.0.0/pbmc_1k_v3/pbmc_1k_v3_filtered_feature_bc_matrix.h5}{official 10x Genomics website}. Using the Julia command \texttt{SparseMatrixCSC}, we processed the sparse matrix into a standard count matrix. The cell annotation file was sourced from \texttt{cluster.csv}, located in the \texttt{analysis/clustering/graphclust/} path within the \href{https://cf.10xgenomics.com/samples/cell-exp/3.0.0/pbmc_1k_v3/pbmc_1k_v3_analysis.tar.gz}{\texttt{pbmc\_1k\_v3\_analysis.tar}} file. For analysis, 151 cells were specifically selected from Cluster 6 of peripheral blood mononuclear cells (PBMCs) derived from a healthy human donor.
\item 
\textbf{SCRB-seq:} The data file \texttt{41564\_2018\_330\_MOESM3\_ESM.xlsx} was obtained from the supplementary materials of Ref. \cite{saint2019single}. In this Excel file, \texttt{Table\_S4} contains the count data, while \texttt{Table\_S3} provides cell annotation information. From this file, we specifically chose the 96 fission yeast cells processed under the experimental condition "37C\_15min".
\item
\textbf{MERFISH:} The count data for U-2 OS cells (human) was obtained from the file \href{https://www.pnas.org/doi/suppl/10.1073/pnas.1912459116/suppl_file/pnas.1912459116.sd12.csv}{\texttt{pnas.1912459116.s\\d12.csv}}, as described in Ref. \cite{xia2019spatial}. From this dataset, we analyzed the count data from 323 cells in Batch 3.
\item
\textbf{FLASH-seq:} The count data (\texttt{HEK\_FS\_lowAmplification\_250K.rds}) for 206 HEK293T cells was obtained from \href{https://data.mendeley.com/datasets/bh47n6fnpd/1}{Mendeley data}, as described in Ref. \cite{hahaut2022fast}. The count matrix was extracted using the R language (provided as \texttt{HEK\_FS\_lowAmplification\_250K.csv}).
\item 
\textbf{Birds in North America:} The data file \texttt{PFW\_all\_2021\_2023\_June2023\_Public.csv} was accessed from \href{https://clo-pfw-prod.s3.us-west-2.amazonaws.com/data/202306/PFW_all_2021_2023_June2023_Public.zip}{Feederwatch}. The data collection protocol was reported in Ref. \cite{bonter2021over} and we focused specifically on observation data from 2022. This data was structured into a count matrix, with each row representing an observation spot and each column representing a bird species.
\item
\textbf{High-energy physics paper citations:} Citation data of high-energy physics papers submitted to arXiv from January 1993 to April 2003 (124 months) were obtained from the files \texttt{Cit-HepTh.txt} and \texttt{Cit-HepTh-dates.txt} on the \href{https://snap.stanford.edu/data/cit-HepTh.html}{Stanford Portal}. The data was filtered to focus on 3444 papers submitted from January to December 1995. The number of citations to this group of papers from papers submitted from January 1996 to December 2002 (84 months) was computed. This citation data was then organized into a count matrix, with each row corresponding to a paper and each column representing a monthly time window.
\item
\textbf{Colorado butterfly:} The data file \texttt{Boulder-Abundance.txt} was downloaded from \href{https://figshare.com/articles/dataset/Data_Paper_Data_Paper/3526085?backTo=/collections/BOULDER_COUNTY_OPEN_SPACE_BUTTERFLY_DIVERSITY_AND_ABUNDANCE/3299114}{Figshare}, the supplementary data of Ref. \cite{oliver2006boulder}. The data for the number of butterflies of a certain species observed at a particular observation site was organised into a count matrix, with each column corresponding to a species and each row representing a different site.
\item
\textbf{E-commerce data in UK:} Data was obtained from \href{https://archive.ics.uci.edu/dataset/352/online+retail}{UCI Machine Learning Repository}, as described in Ref. \cite{chen2012data}. The data was filtered to focus solely on customer purchases from the United Kingdom. This filtered dataset was then transformed into a standard count matrix, where the value in the $i$-th row and $j$-th column represents the number of item $i$ purchased by customer $j$. Data is provided in the file \texttt{e-comm.csv}.
\item 
\textbf{Disease outbreak in US:} Weekly data was collected from the website for the \href{https://wonder.cdc.gov/nndss/nndss_weekly_tables_menu.asp?mmwr_year=2024&mmwr_week=12&comingfrom=202413&savedmode=}{Centers for Disease Control and Prevention (CDC) in the United States} from 2022 to 2024. Specifically we focused on data from the \href{https://wonder.cdc.gov/nndss/nndss_weekly_tables_menu.asp?comingfrom=202412&savedmode=&mmwr_year=2024&mmwr_week=10}{10th week of 2024}. The data for the number of people with a particular disease in a particular state was organised into a count matrix, with each column corresponding to a state and each row representing an infectious disease or condition. The preprocessed data is stored as \texttt{cdc.csv}.
\item
\textbf{LinkedIn job in US:} The employment data was obtained for locations in the USA from the file \texttt{postings.csv} on \href{https://www.kaggle.com/datasets/arshkon/linkedin-job-postings}{Kaggle}. A count matrix was structured with columns representing the 50 states, 1 federal district, and 5 inhabited territories (a total of 56 regions, provided as the file \texttt{us-states-terr\\itories.csv}) and rows representing different occupations.

\end{itemize}

\section{Region with no points in the mean-Fano factor plots} \label{nopoints}

Consider a gene whose counts are described by some arbitrary discrete distribution $P(n)$ with mean $\langle n \rangle$ and variance $\sigma^2$. From the definition of the variance of the counts, it follows that
\begin{align}
    \sigma^2 &= \langle n^2 \rangle - \langle n \rangle^2 \ge \langle n \rangle - \langle n \rangle^2, \notag \\
    {\rm{FF}} &= \frac{\sigma^2}{\langle n \rangle} \ge 1 - \langle n \rangle, \notag \\
    \langle n \rangle &\ge 1 - {\rm{FF}}.
\end{align}
Note that in the first line we have used that $n^2 \ge n$ for any positive integer $n$. Hence it follows that the region $\langle n \rangle < 1 - {\rm{FF}}$ in the mean-Fano factor plot is devoid of points.

\section{General theory for patterns in mean-Fano factor plots} \label{gencurve}

Consider genes which can have only 0, 1, 2, $\cdots$, $N$ transcripts per cell with the proviso that there are exactly $m_i$ cells which have exactly $i$ transcripts for $2 \le i \le N$. Note that we do not fix the number of cells with 0 or 1 transcripts, i.e. $m_0$ and $m_1$ can vary between different genes.  
Let the total number of cells be $\sum_{i=0}^N m_i = n_c$. It then follows that the mean and the mean squared expression are given by:
\begin{align}
    \label{meanexpgen}
    \langle n \rangle &= \frac{\sum_{i=1}^N i m_i }{n_c} = a + \frac{\sum_{i=2}^N i m_i}{n_c},   \\
    \langle n^2 \rangle &= \frac{\sum_{i=1}^N i^2 m_i }{n_c} = a + \frac{\sum_{i=2}^N i^2 m_i}{n_c},
\end{align}
where $m_1 /n_c$ equals some fraction $a$ (it can only take values $0$, $1/n_c$, $\cdots$, $1 - \sum_{i=2}^N m_i/n_c$). Hence the Fano factor is given by
\begin{equation}
    {\rm{FF}}= \frac{ \langle n^2 \rangle -  \langle n \rangle^2}{\langle n \rangle} = 1 - \langle n \rangle + \frac{\sum_{i=2}^N i(i-1) m_i}{n_c \langle n \rangle},
    \label{FFgen}
\end{equation}
where we eliminated the parameter $a$ using Eq. \eqref{meanexpgen}. Note that Eqs. \eqref{FFgen} is independent of $m_0$ and $m_1$ and hence it is valid independent of the number of cells with 0 or 1 transcripts. 

Hence it follows that for a fixed cell number vector $\vec{v} = \{m_2, \cdots, m_N\}$, Eq. \eqref{meanexpgen} gives the y-coordinate and Eq. \eqref{FFgen} gives the x-coordinate of $1 - \sum_{i=2}^N m_i/n_c$ points in the mean-Fano factor plot, where each point corresponds to one or more genes. If we consider all possible values of the entries of the vector $\vec{v}$ and vary $N$ as well then all of the mean-FF plot space is scanned. Since $i(i-1)$ is an even number for $ i\ge 2$, it follows from Eq. \eqref{FFgen} that the  x-coordinate of a general point is given by Eq. \eqref{FFeq} but with the difference that that the curve number $k$ varies between $1$ and $\infty$. The y-coordinates are multiples of $1/n_c$ in the range $1/n_c$ to $\infty$.
\vspace*{6cm}

\aboverulesep=0ex
 \belowrulesep=0ex
\begin{table} [h]
\centering
\caption{Medians of $m_0/n_c$ and $m_1/n_c$ grouped by dataset and theoretical curve number $k$}\label{tabS1}\vspace*{1em}
\resizebox{0.8\textwidth}{!}{
\begin{tabular}{c|ccc|ccc}
\toprule
&\multicolumn{3}{c|}{$m_0/n_c$}&\multicolumn{3}{c}{$m_1/n_c$}\\
\cmidrule(l{0pt}r{0pt}){2-7}
& Curve 0 & Curve 1 & Curve 2 & Curve 0 & Curve 1 & Curve  2 \\
\hline
Smartseq-3& 0.902 & 0.895 &  0.871 & 0.098 &   0.098 & 0.115 \\
VASA-seq& 0.881 & 0.871 & 0.841 & 0.119 & 0.119 & 0.139 \\
10x Genomics v3& 0.907 & 0.901 & 0.877 & 0.093 & 0.093 & 0.109\\
SCRB-seq& 0.875 & 0.844 & 0.823 & 0.125 & 0.146 & 0.156  \\
MERFISH& 0.916 & 0.898 & 0.838 & 0.084 & 0.099 & 0.156 \\
FLASH-seq& 0.932 & 0.922 & 0.910 & 0.068 & 0.073 & 0.080 \\
\bottomrule
\end{tabular}
}
\end{table}

\end{document}